\documentclass[twocolumn]{aastex63}

\usepackage{times}
\usepackage{amsmath}
\usepackage{graphicx}
\usepackage{subfigure}
\usepackage{placeins} 
\usepackage{hyperref}
\usepackage{gensymb}
\usepackage{upgreek}
\def\H2{$\rm H_{2}$}
\newcommand{\MHI}{$M_{\mathrm{H\, \textsc{i}}}$\,}
\newcommand{\HI}{${\mathrm{H\, \textsc{i}}}$\,}

\newcommand{\tHI}{$\tau_{\mathrm{H\, \textsc{i}}}$\,}

\begin{document}

\title{\large{Spiral Structure Boosts Star Formation in Disk Galaxies}}

\author[0000-0002-3462-4175]{Si-Yue Yu}
\affiliation{Kavli Institute for Astronomy and Astrophysics, Peking University, Beijing 100871, China}
\affiliation{Department of Astronomy, School of Physics, Peking University, Beijing 100871, China}

\author[0000-0001-6947-5846]{Luis C. Ho}
\affiliation{Kavli Institute for Astronomy and Astrophysics, Peking University, Beijing 100871, China}
\affiliation{Department of Astronomy, School of Physics, Peking University, Beijing 100871, China}

\author[0000-0002-6593-8820]{Jing Wang}
\affiliation{Kavli Institute for Astronomy and Astrophysics, Peking University, Beijing 100871, China}
\affiliation{Department of Astronomy, School of Physics, Peking University, Beijing 100871, China}

\correspondingauthor{Si-Yue Yu}
\email{syu@mpifr-bonn.mpg.de}

\begin{abstract}
We investigate the impact of spiral structure on global star formation using a sample of 2226 nearby bright disk galaxies. Examining the relationship between spiral arms, star formation rate (SFR), and stellar mass, we find that arm strength correlates well with the variation of SFR as a function of stellar mass.  Arms are stronger above the star-forming galaxy main sequence (MS) and weaker below it: arm strength increases with higher $\log\,({\rm SFR}/{\rm SFR}_{\rm MS})$, where ${\rm SFR}_{\rm MS}$ is the SFR along the MS. Likewise, stronger arms are associated with higher specific SFR. We confirm this trend using the optical colors of a larger sample of 4378 disk galaxies, whose position on the blue cloud also depends systematically on spiral arm strength. This link is independent of other galaxy structural parameters. For the subset of galaxies with cold gas measurements, arm strength positively correlates with \HI\ and \H2\ mass fraction, even after removing the mutual dependence on $\log\,({\rm SFR}/{\rm SFR}_{\rm MS})$, consistent with the notion that spiral arms are maintained by dynamical cooling provided by gas damping. For a given gas fraction, stronger arms lead to higher $\log\,({\rm SFR}/{\rm SFR}_{\rm MS})$, resulting in a trend of increasing arm strength with shorter gas depletion time.  We suggest a physical picture in which the dissipation process provided by gas damping maintains spiral structure, which, in turn, boosts the star formation efficiency of the gas reservoir. 
\end{abstract}

\section{Introduction}

The manner in which stars form and how the star formation process is regulated in galaxies are major themes in extragalactic astronomy.  Stellar mass and star formation rate (SFR) in star-forming disk galaxies display a surprising degree of regularity, which is often expressed as the galaxy main sequence (MS) observed in both the local \citep[e.g.,][]{Salim2007, Renzini2015} and high-redshift \citep[e.g.,][]{Daddi2007, Noeske2007, Whitaker2014} Universe. However, this scaling relation contains significant intrinsic scatter, and much effort has been invested to understand what physical drivers control the scatter.  The distance of a galaxy from the MS, or specific SFR (sSFR), correlates with the galaxy's gas content and star formation efficiency (SFE, defined as SFR per unit gas mass): galaxies above the MS are associated with increased gas fraction \citep{Saintonge2012, Saintonge2016, Genzel2015, Wang2020} and higher SFE \citep{ Daddi2010, Genzel2010,Saintonge2011b, Genzel2015, Saintonge2016}. It has also been established that most of the galaxies that lie well above the MS are major mergers \citep{Sanders1996, Veilleux2002, Engel2010}. In contrast, galactic stellar bars only have a marginal effect on elevating SFE \citep{Saintonge2012}.

Spiral arms---one of the most generic organized substructures in disk galaxies---may influence how stars form. Theoretical models show that spiral arms can induce large-scale shocks on gas clouds as they pass through a spiral arm, trigger gravitational collapse, and accelerate the production of new stars \citep{Roberts1969}. By employing arm concentration and arm amplitude to measure spiral arm strength, \cite{Seigar2002} and \cite{Kendall2015} report that stronger arms trigger more significant shocks, which enhance SFR or sSFR. On the other hand, dynamical cooling provided by gas damping is required to maintain spiral structure. Without cold gas, galactic disks would continue to heat up until they become stable to density wave perturbation \citep{Binney} or local gravitational instability \citep{Sellwood1984}.  Spiral arms may play an intertwined role with gas content and gas compression, and hence with the overall star formation process. Studying the interdependence of spiral structure on global star formation sheds light not only on the star formation process but also on how spiral arms evolve in disk galaxies.

\cite{YuHo2020} analyze the spiral structure of a large, comprehensive sample of nearby galaxies.  We use their catalog of spiral arm strength to study the impact of spiral structure on the global star formation process of disk galaxies.

\section{Observational Material}
\subsection{Data}
This study is based on \cite{YuHo2020}, who quantify spiral structure in a large, comprehensive sample of 4378 nearby (redshift $z \le 0.03$), bright ({\it r}-band magnitude $\le14.5$), face-on (inclination angle $\le 60 \degree$) disk (spiral and lenticular) galaxies selected from the Sloan Digital Sky Survey \citep[SDSS;][]{York2000} using the NASA Sloan Atlas (NSA) catalog \citep{Blanton2011}\footnote{\url{http://www.nsatlas.org/}}. The selection criteria are based on extensive simulations \citep{Yu2018a} to ensure robust measurement of spiral arm properties using images similar in character to those from SDSS. To take advantage of published \HI\ and CO measurements, in this disk sample we slightly relax the restriction on $z$ and $m_r$ to include galaxies acquired from the extended GALEX Arecibo SDSS survey \citep[xGASS;][]{Catinella2018} and the extended CO Legacy Database for GASS \citep[xCOLD GASS;][]{Saintonge2017}.

We utilize estimates of SFR and stellar mass ($M_*$) based on ultraviolet, optical, and mid-infrared photometry from the xGASS catalog for galaxies contained therein, and from the catalog of \cite{Salim2018}\footnote{\url{http://pages.iu.edu/~salims/gswlc/}; we use GSWLC-X.} for the rest. This results in 2226 (1751 spiral and 475 lenticular) galaxies, the main sample studied in this work.  For the purposes of securing a larger sample to generate Figure~2b, we use $M_*$ from the NSA catalog, after adjusting the values by an offset of 0.27\,dex to be consistent with \cite{Salim2018}.  The rest-frame, extinction-corrected absolute magnitudes are from the NSA catalog. The MS \citep{Saintonge2016} follows $\log {\rm SFR_{\rm MS}}=-2.332x + 0.4156x^2-0.01828x^3$, where $x = \log(M_*/M_{\odot})$, and the distance from the MS is defined as $\log\,({\rm SFR}/{\rm SFR}_{\rm MS})$, where SFR$_{\rm MS}$ is the SFR along the MS. The measurement uncertainty of SFR and $M_*$ will not affect our results (Section~3.5). 
In the main sample, the \HI\ masses of 356 galaxies were derived from integrated \HI\ spectra from xGASS, and the H$_2$ masses of 141 galaxies were based on CO(1--0) observations from xCOLD GASS.  Non-detections and confused sources that have more than one potential galaxies contributing to the HI spectrum have been excluded. The sample covers a wide range of $M_*$ ($\sim 10^{8.8} - 10^{11.4}\, M_\odot$) and SFR ($\sim 0.001 - 10\,M_\odot\,{\rm yr}^{-1}$). All distances are based on a $\Lambda$CDM cosmology with $\Omega_m=0.3$, $\Omega_{\lambda}=0.7$, and $h=0.7$.

\subsection{Spiral Arm Strength}

\cite{YuHo2020} extract the {\it r}-band galaxy light profile as a function of azimuthal angle along an ellipse of fixed ellipticity and position angle at each radius. Fitting a Fourier series to the profile, they define the arm amplitude ($s_{\rm arm}$) as the average Fourier amplitude (quadratic sum of $m=2$, $3$, and $4$ modes) relative to an axisymmetric disk ($m=0$ mode) over a radial range that starts at the end of the bulge or bar and ends at $R_{90}$, the radius that encloses 90\% of the total flux and encompasses the majority of the spiral structure \citep{Yu2018a}.  Due to the highly nonlinear connection between spiral arms with star formation and gas fraction, as explored in this work, we use the logarithmic form ($\log\,s_{\rm arm}$) to denote spiral arm strength.

The noise from sky background will drown out the signal of faint structure,  especially in the outskirt of the galaxy, when the signal-to-noise ratio (SNR) is sufficiently low.  It is expected that the Fourier amplitude and hence the spiral arm strength will be overestimated because the noise will also contribute to the Fourier decomposition. Therefore, it is necessary to quantify and correct the systematic bias of spiral arm strength caused by noise. The key quantity to control the bias is the average SNR of each pixel, instead of the SNR of the total galaxy flux, because our analysis is based on the intensity distribution along isophotes at each radius. To estimate the bias, we first define the SNR as the ratio of average pixel value over the radial region from $R_{50}$ to $R_{90}$ to the sky Poisson noise. The inner boundary $R_{50}$ is set to avoid the central region. 90\%, 75\%, and 48\% of galaxies in our sample have SNR $>2$, $3$, and $5$, respectively. We then select 100 galaxies with SNR ranging from 15 to 30, which are sufficient high to quantify spiral arm strength ($\log\,s_{\rm arm}$) without bias.  We then add Poisson noise to the 100 galaxies to generate noisier galaxies with the target SNR, and we apply the same measurement method to these noisy galaxies to obtain the spiral arm strength with bias ($\log\,s^{\rm noisy}_{\rm arm}$). The systematic bias is calculated as $\Delta(\log\,s_{\rm arm}) = \log\,s_{\rm arm}-\log\,s^{\rm noisy}_{\rm arm}$. As illustrated in Figure~1, the systematic bias caused by noise is small, and it is corrected for our measured arm strength. Larger $\log\,s_{\rm arm}$ corresponds to stronger spiral arms (see examples on the right of Figure~2).

To facilitate our investigation, we use the results of \cite{YuHo2020} to calculate the bar strength ($s_{\rm bar}$) as the average $m=2$ Fourier amplitude over the bar region. The bar, locating in the center of galaxy, has high SNR, and hence no bias caused by noise is calculated. We calculate galaxy concentration index as $C = 5 \log (R_{80}/R_{20}$), where $R_{80}$ and $R_{20}$ are the radii that enclose 80\% and 20\% of the light, respectively. We calculate stellar surface density ($\mu_*$) by dividing the stellar mass by the circular area with {\it r}-band half-light radius.

\begin{figure}
\begin{center}
\includegraphics[width=0.45\textwidth]{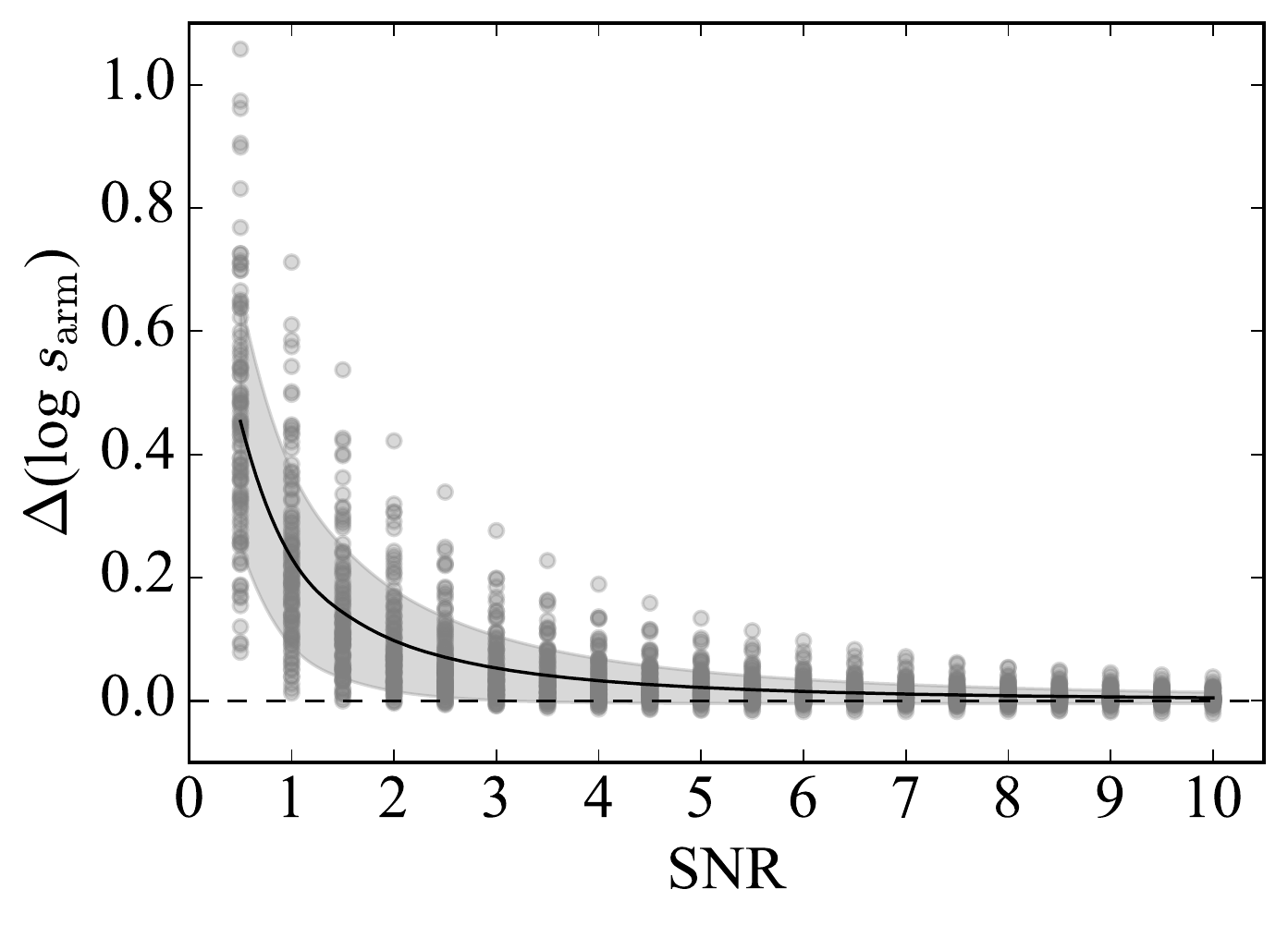}
\caption{Systematic bias of spiral arm strength ($\log\,s_{\rm arm}$) as a function of signal-to-noise ratio (SNR). The solid curve and shaded area mark the mean and scatter for a given SNR.
} 
\label{default}
\end{center}
\end{figure}

%
\begin{figure*}[t]
\begin{center}
\includegraphics[width=\textwidth]{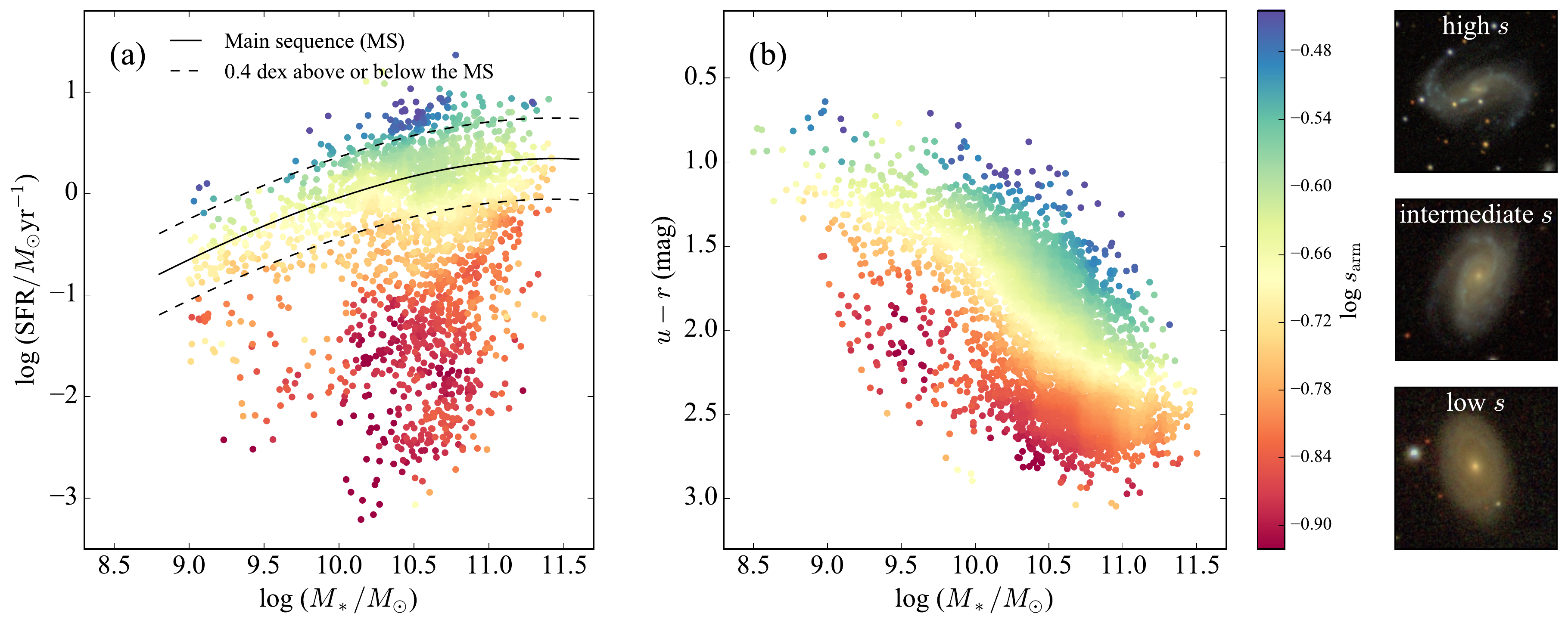}
\caption{
The influence of spiral arms on star formation in disk galaxies. Panel (a) illustrates the spiral arm strength ($\log\,s_{\rm arm}$) as a function of SFR and $M_*$ for 2226 face-on disk galaxies with available SFRs based on multi-band photometry. The solid curve marks the MS \citep{Saintonge2016}, with the dashed curves indicating 0.4\,dex above and below it.  Panel (b) gives the spiral arm strength as a function of rest-frame, extinction-corrected $u-r$ color and $M_*$ for the full sample of 4378 disk galaxies.  The color associated with each data point encodes the average arm strength of surrounding galaxies with $\Delta x \le 0.2$\,dex and $\Delta y \le 0.2$\,dex.  The images on the right show example galaxies with spirals of strong, intermediate, and weak arm strength.
}
\label{default}
\end{center}
\end{figure*}

\begin{figure*}
\begin{center}
\plotone{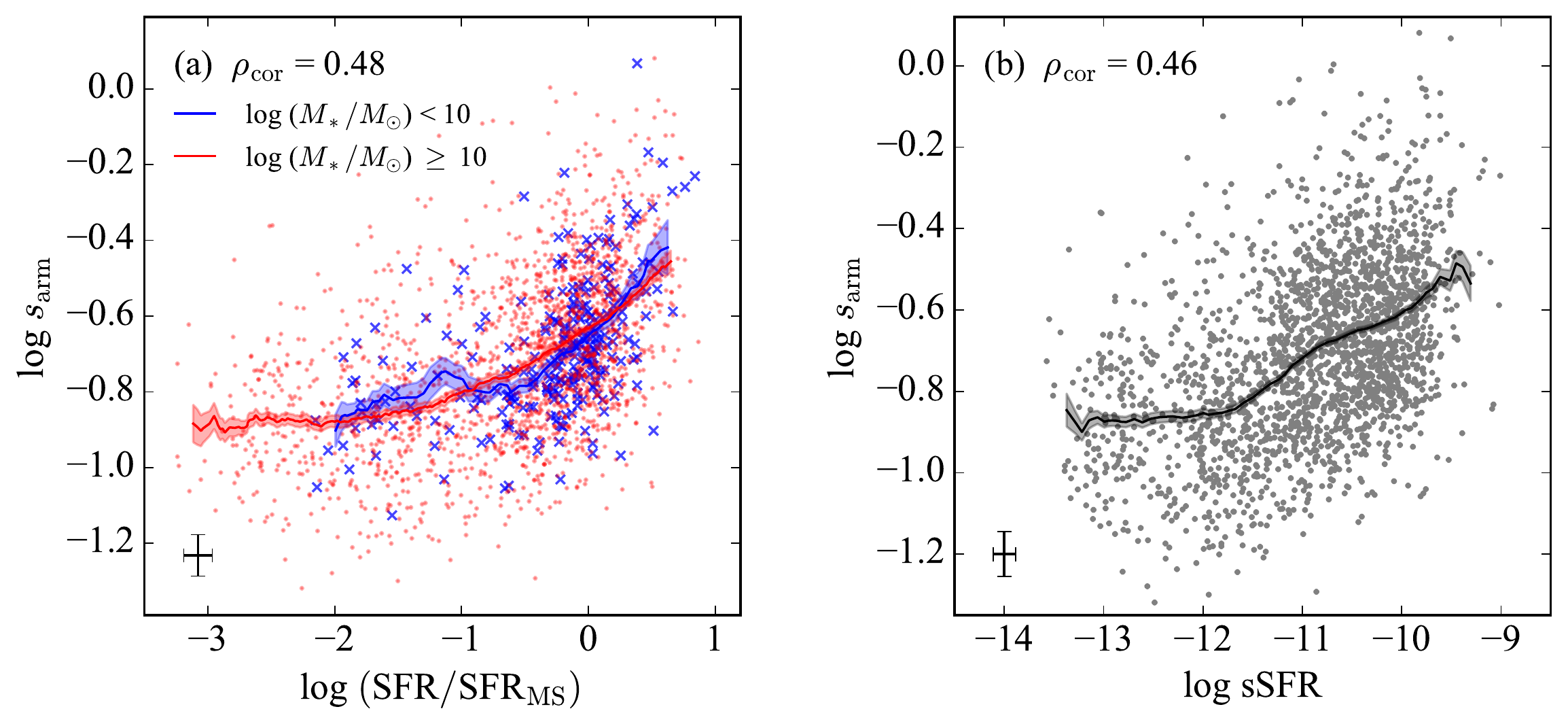}
\caption{
The relationship between spiral arm strength ($\log\,s_{\rm arm}$) and (a) distance from the MS [$\log\,({\rm SFR}/{\rm SFR}_{\rm MS})$] and (b) specific SFR (sSFR). In panel (a), the blue crosses and red points denote galaxies with $\log\,(M_*/M_{\odot})<10$ and $\log\,(M_*/M_{\odot})\ge 10$, respectively.  Solid curves mark the mean value, while shaded regions mark the error of the mean value. The bar on the lower-left corner indicates the typical measurement uncertainty. The correlation coefficient $\rho_{\rm cor}$ is given in the upper-left corner of each panel.}
\label{default}
\end{center}
\end{figure*}

\section{Results and Discussion}

\subsection{The Influence of Spiral Arms on  the MS}

Figure~2a displays the familiar dependence of SFR on galaxy $M_*$, which predominantly manifests itself as a star-forming MS \citep[solid curve;][]{Saintonge2016}, and a long tail of low SFR for quenched galaxies deviating from the MS. We reveal an effect not previously seen: at fixed $M_*$, the location of a galaxy on the SFR-$M_*$ plane correlates well with the strength of its spiral arms. Spiral arms are on average stronger above the MS and weaker below it. We confirm this trend for the full sample of 4378 disk galaxies by replacing SFR with the rest-frame, extinction-corrected $u-r$ color index (Figure~2b). Within the blue cloud \citep{Strateva2001, Blanton2003}, spiral arms are on average stronger in bluer, younger galaxies, becoming progressively weaker through the green valley and into the red sequence of quenched disk galaxies.  The quenched disk galaxies are lenticulars or red spirals with little or no spiral features.

\begin{figure*}
\begin{center}

\includegraphics[width=1.0\textwidth]{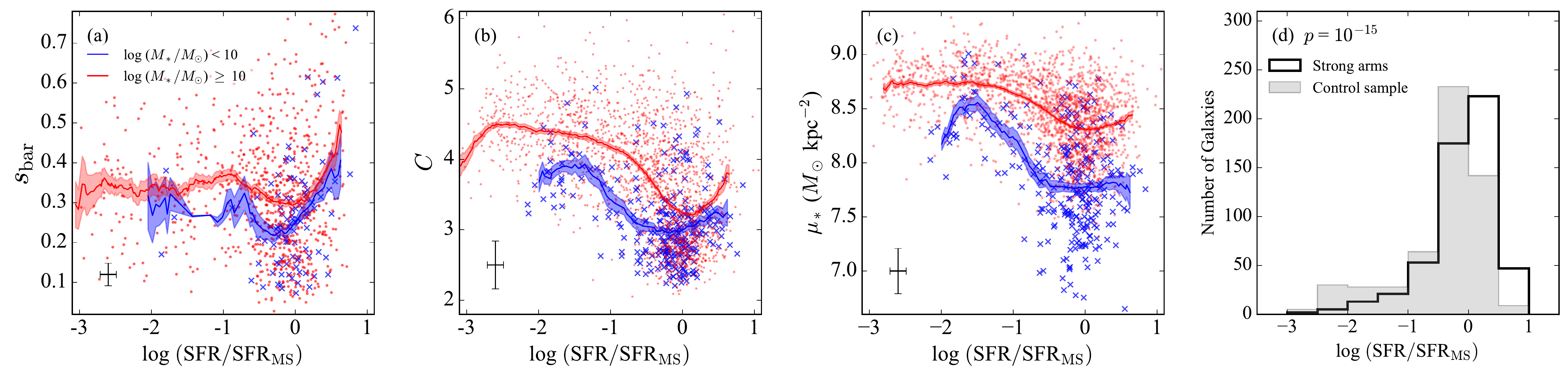}
\caption{
The effect of (a) bar strength ($s_{\rm bar}$), (b) concentration index ($C$), and (c) stellar mass surface density ($\mu_*$) on the distance from the MS [$\log\,({\rm SFR}/{\rm SFR}_{\rm MS})$].  The blue crosses and red points denote galaxies with $\log\,(M_*/M_{\odot})<10$ and $\log\,(M_*/M_{\odot}) \ge 10$, respectively. Solid curves mark the average value, while shaded regions mark the error of the mean value. The bar on the lower-left corner indicates the typical measurement uncertainty. Panel (d) shows the distribution of $\log\,({\rm SFR}/{\rm SFR}_{\rm MS})$ for the strong-armed sample ($\log s_{\rm arm} \ge  -0.57$) and the weak-armed ($\log s_{\rm arm} < -0.57$) control sample, which has similar $M_*$, $C$, and $\mu_*$ as the strong-armed sample.
}
\label{default}
\end{center}
\end{figure*}

\begin{figure*}
\begin{center}
\includegraphics[width=0.7\textwidth]{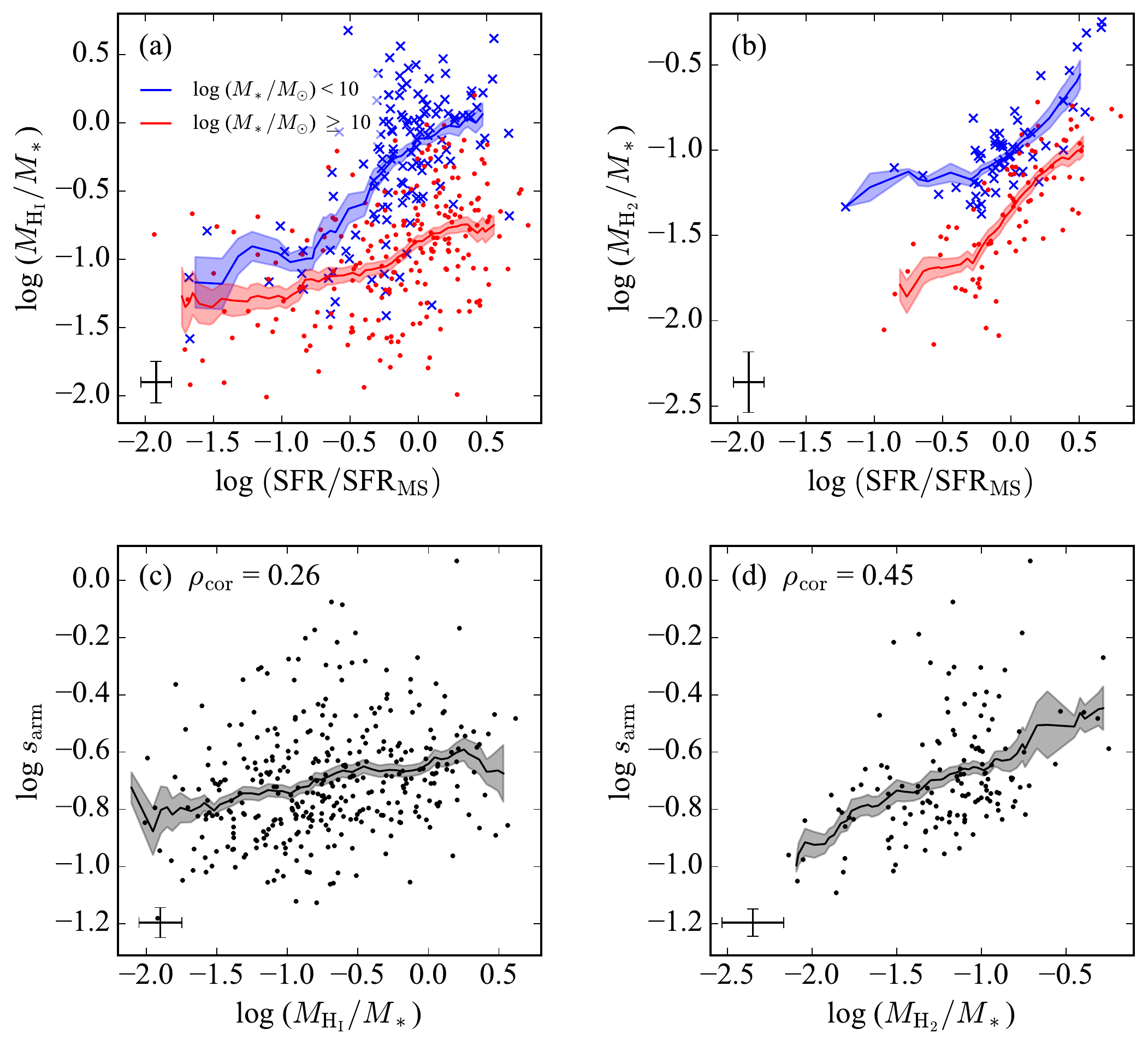}
\caption{
The relationship between atomic (\MHI$/M_*$) and molecular ($M_{\rm H_2}/M_*$) gas fraction, distance from the MS [$\log\,({\rm SFR}/{\rm SFR}_{\rm MS})$], and spiral arm strength ($\log\,s_{\rm arm}$).  Panels (a) and (b) show the variation of $\log\,({\rm SFR}/{\rm SFR}_{\rm MS})$ with \MHI$/M_*$ and $M_{\rm H_2}/M_*$. The blue crosses and red points denote the galaxies with $\log\,(M_*/M_{\odot})<10$ and $\log\,(M_*/M_{\odot})\ge10$. Panels (c) and (d) present the dependence of spiral arm strength ($\log\,s_{\rm arm}$) on \MHI$/M_*$ and $M_{\rm H_2}/M_*$; the correlation coefficient $\rho_{\rm cor}$ is given.  Solid curves mark the mean value, while shaded regions mark the error of the mean value. The bars indicate typical measurement uncertainties.}
\label{default}
\end{center}
\end{figure*}

\begin{deluxetable}{ccccc}

\setlength\tabcolsep{3pt}

\tablenum{1}
\tablecaption{Partial Correlation Analysis}
\tablehead{
\colhead{Parameter 1} &
\colhead{Parameter 2} &
\colhead{Dependence} &
\colhead{Partial}  &
\colhead{$p$ value} \\
\colhead{} &
\colhead{} &
\colhead{Removed} &
\colhead{Corr. Coeff.}
}
\startdata
$\log\,s_{\rm arm}$  & $\log\,({\rm SFR}/{\rm SFR}_{\rm MS})$  & $\dots$  &  $0.48$ & $<10^{-3}$\\
$\log\,s_{\rm arm}$  & $\log\,({\rm SFR}/{\rm SFR}_{\rm MS})$  & $s_{\rm bar}$  &  $0.55$ & $<10^{-3}$\\
$\log\,s_{\rm arm}$  & $\log\,({\rm SFR}/{\rm SFR}_{\rm MS})$  & $C$  &  $0.41$ & $<10^{-3}$ \\
$\log\,s_{\rm arm}$  & $\log\,({\rm SFR}/{\rm SFR}_{\rm MS})$  & $\log\,\mu_*$  &  $0.39$ & $<10^{-3}$\\
\hline
$\log\,s_{\rm arm}$  & \MHI$/M_*$  & $\dots$  &  $0.26$ & $<10^{-3}$\\
$\log\,s_{\rm arm}$  & \MHI$/M_*$  & $\log\,({\rm SFR}/{\rm SFR}_{\rm MS})$  &  $0.10$ & 0.065\\
$\log\,s_{\rm arm}$  & $\log\,({\rm SFR}/{\rm SFR}_{\rm MS})$ & \MHI$/M_*$ &  $0.41$ & $<10^{-3}$\\
$\log\,s_{\rm arm}$  & \tHI\ &   $\dots$ & $-0.21$ & $<10^{-3}$\\
$\log\,s_{\rm arm}$  & \tHI\ &   $\log\,\mu_{*}$ & $-0.25$ & $<10^{-3}$\\
\hline
$\log\,s_{\rm arm}$  & $M_{\rm H_2}/M_*$  & $\dots$  &  $0.45$ & $<10^{-3}$\\
$\log\,s_{\rm arm}$  & $M_{\rm H_2}/M_*$  & $\log\,({\rm SFR}/{\rm SFR}_{\rm MS})$  &  $0.20$ & 0.017\\
$\log\,s_{\rm arm}$  & $\log\,({\rm SFR}/{\rm SFR}_{\rm MS})$ & $M_{\rm H_2}/M_*$ &  $0.33$ & $<10^{-3}$\\
$\log\,s_{\rm arm}$  & $\tau_{\rm H_2}$ &   $\dots$ & $-0.30$ & $<10^{-3}$\\
$\log\,s_{\rm arm}$  & $\tau_{\rm H_2}$ &   $\log\,\mu_{*}$ & $-0.22$ & 0.009\\
\enddata
\end{deluxetable}

The spiral arm strength is unambiguously linked to the global, galaxy-scale star formation. Figure~3a illustrates the correlation between arm strength and the distance from the MS, $\log\,({\rm SFR}/{\rm SFR}_{\rm MS})$. Average spiral arm strength rises markedly and monotonically with $\log\,({\rm SFR}/{\rm SFR}_{\rm MS})$, despite the flattening far below the MS due to the lack of spirals. The correlation coefficient is $\rho_{\rm cor}=0.48$ with $p$ value $<0.001$ (Table~1).  This correlation has no obvious dependence on stellar mass, as illustrated by the high-mass ($\log(M_*/M_{\odot})>10$; red) and low-mass ($\log(M_*/M_{\odot})<10$; blue) subsample. The specific SFR (sSFR) is another indicator of distance from the MS \citep{Saintonge2011b}.  Likewise, Figure~3b presents a similar relationship between arm strength and sSFR. The trends are consistent with the studies by \cite{Seigar2002} and \cite{Kendall2015}, even though their samples, with only about a dozen objects, are severely limited by small-number statistics.  Our work is established at a much more statistically significant level.  In the rest of this work, we will use $\log\,({\rm SFR}/{\rm SFR}_{\rm MS})$ as the main metric to evaluate the level of star formation in a galaxy.  The results based on sSFR are similar. 

No other structural feature in the galaxy imparts as clear-cut an influence on the MS as do spiral arms.  While bars impose a non-axisymmetric gravitational potential on the disk and may drive spiral structure \citep{Buta2009, Durbala2009}, we find that bars, whose strength we characterize as $s_{\rm bar}$, have no obvious influence on $\log\,({\rm SFR}/{\rm SFR}_{\rm MS})$, except for the galaxies above the MS (Figure~4a).  In any case, the half of the galaxy population that lack bars presents the same relation between arm strength and $\log\,({\rm SFR}/{\rm SFR}_{\rm MS})$.  A classical bulge is dynamically stable against the formation of substructures, and hence may suppress spiral arms by reducing the mass fraction of the dynamically active disk that reacts to spiral perturbation \citep{Bertin1989}. Using the concentration index ($C$) as a proxy to gauge the degree of bulge prominence, Figure~4b reveals no monotonic correlation between $C$ and $\log\,({\rm SFR}/{\rm SFR}_{\rm MS})$, apart from the known U-shaped trend for $C$ to increase above and below the MS \citep{Bluck2019, Popesso2019}. Stellar mass plays a role in that $C$ decreases for less massive galaxies for a given $\log\,({\rm SFR}/{\rm SFR}_{\rm MS})$. Lastly, we examine the role of stellar mass surface density within the half-light radius ($\mu_*$), which has been suggested to be associated with star formation quenching \citep{Franx2008, Whitaker2017}. We find that $\mu_*$ behaves quite similarly to $C$, especially in its dependence on stellar mass (Figure~4c).  The partial correlation coefficient for the residual dependence of arm strength on $\log\,({\rm SFR}/{\rm SFR}_{\rm MS})$, after removing the mutual dependence on $s_{\rm bar}$, $C$, and $\mu_*$, remains strong and statistically significant: $\rho_{\rm cor} = 0.55$, 0.41, and 0.39, respectively, with $p$ values $<0.001$ (Table~1).

It is also instructive to test the joint effect of $M_*$, $\mu_*$, and $C$. We construct a strong-armed sample using the subset of galaxies that have spiral arm strengths that occupy the top 20\% of the full sample ($\log s_{\rm arm} \ge-0.57$).  For each galaxy in the strong-armed sample, we construct a control sample by randomly assigning a relatively weak-armed ($\log s_{\rm arm} <-0.57$) galaxy with similar stellar mass ($\Delta \log M_*<0.1$), stellar surface density ($\Delta \log \mu_*<0.1$), and concentration ($\Delta C<0.1$). As illustrated in Figure~4d, the strong-armed sample has higher $\log\,({\rm SFR}/{\rm SFR}_{\rm MS})$ than the control sample at a significant confidence level.  The Kolmogorov-Smirnov test rejects the null hypothesis that the two samples are drawn from the same parent distribution with a probability of $p=10^{-15}$. Therefore, neither the strength of the bar, the prominence of the bulge, nor the stellar surface density can explain the connection between arm strength and the position of spiral galaxies on the SFR-$M_*$ plane or the blue cloud.

\subsection{The Role of Gas}

To gain deeper insight into the physical origin of the correlation between arm strength and $\log\,({\rm SFR}/{\rm SFR}_{\rm MS})$, we must consider the role of gas in regulating both star formation and spiral structure. The gas fraction is regarded as one of the main drivers of a galaxy's position on the SFR-$M_*$ diagram \citep{Saintonge2016}.  As with the spiral arm strength, $\log\,({\rm SFR}/{\rm SFR}_{\rm MS})$ increases with increasing atomic (\MHI$/M_*$) and molecular ($M_{\rm H_2}/M_*$) gas fraction (Figures~5a and 5b), but the two relations shift systematically toward higher gas fraction for less massive galaxies.  By comparison, the relation between $\log\,({\rm SFR}/{\rm SFR}_{\rm MS})$ and arm strength is nearly identical for galaxies of all stellar masses (Figure~3a).

Figures~5c and 5d reveal a new dependence of spiral arm strength on atomic gas fraction for the subset of 356 galaxies with \HI\ observations ($\rho_{\rm cor}=0.26$ with $p$ value $<0.001$) and on molecular gas fraction for the 141 galaxies with CO measurements ($\rho_{\rm cor}=0.45$ with $p$ value $<0.001$). After removing the mutual dependence on $\log\,({\rm SFR}/{\rm SFR}_{\rm MS})$, the residual dependence of arm strength on atomic gas fraction becomes weak ($\rho_{\rm cor}=0.10$). The $p$ value gives 0.065 ($>0.05$) and hence we cannot reject the null hypothesis that this residual dependence does not exist. The residual dependence for molecular gas fraction is moderate and statistically significant ($\rho_{\rm cor}=0.20$ with $p$ value $=0.017$).  At odds with our results, \cite{Kendall2015} concluded that arm strength correlates inversely with gas fraction, although it is difficult to evaluate the significance of this discrepancy given the very small size of their sample (only 13 objects). Our results are consistent with the expectations of density wave theory \citep{LinShu64} for grand-design spiral galaxies \citep{Elmegreen2011}.  In the model for self-regulated spiral density waves, the gas damps the spiral structure, which is regenerated on timescales comparable to the damping timescale \citep{BertinRomeo1988}. Spiral waves grow exponentially with time in the linear limit. As stronger spirals grow faster, more gas damping is needed to maintain stronger arms. In the absence of gas, the energy in spiral waves transfers into the random motions of stars, which heat the disk until it becomes stable against spiral perturbation by density waves \citep{Binney}.  Gas damping is also important for flocculent spiral galaxies, whose transient arms arise naturally in numerical simulations. A transient spiral can heat up the disk in a few rotations to prevent formation of asymmetric structure \citep{Sellwood1984}. Recent high-resolution {\it N}-body simulation demonstrate that the heating of transient arms is much less than previously thought and the arms can last for a few Gyr, undergoing a cyclical behavior: they are stretched and break under the influence of galactic shear, but self-gravity in locally overdense regions regenerate segments that reconnect into spiral arms \citep{Onghia2013, Fujii2011}.  Nevertheless, without cooling from gas, the simulated galactic disk rapidly becomes more stable with time, while strong arms in an initially unstable disk weaken as the disk stabilizes \citep{Fujii2011}. The evolution cycle of spirals causes the arm strength to fluctuate at the level of $10\%-20\%$ on a timescale of $\sim 1$ Gyr \citep{Fujii2011} and may account for part of the scatter in arm strength at a fixed $\log\,({\rm SFR}/{\rm SFR}_{\rm MS})$.  The decline of arm strength toward the sequence of quenched disks suggests that spiral arms fade away in the absence of cold gas.   The relationship between spiral arms and star formation (Figure~3) is driven only in part by  gas fraction via its role in maintaining spiral structure, because after removing the mutual dependence on \MHI$/M_*$ and $M_{\rm H_2}/M_*$, the residual dependence of arm strength on $\log\,({\rm SFR}/{\rm SFR}_{\rm MS})$ remains strong ($\rho_{\rm cor} = 0.41$ and 0.33, respectively; Table~1) and statistically significant ($p$ values $<0.001$). The partial correlation analysis suggests that spiral arms enhance star formation activity.

\begin{figure*}
\begin{center}
\includegraphics[width=0.8\textwidth]{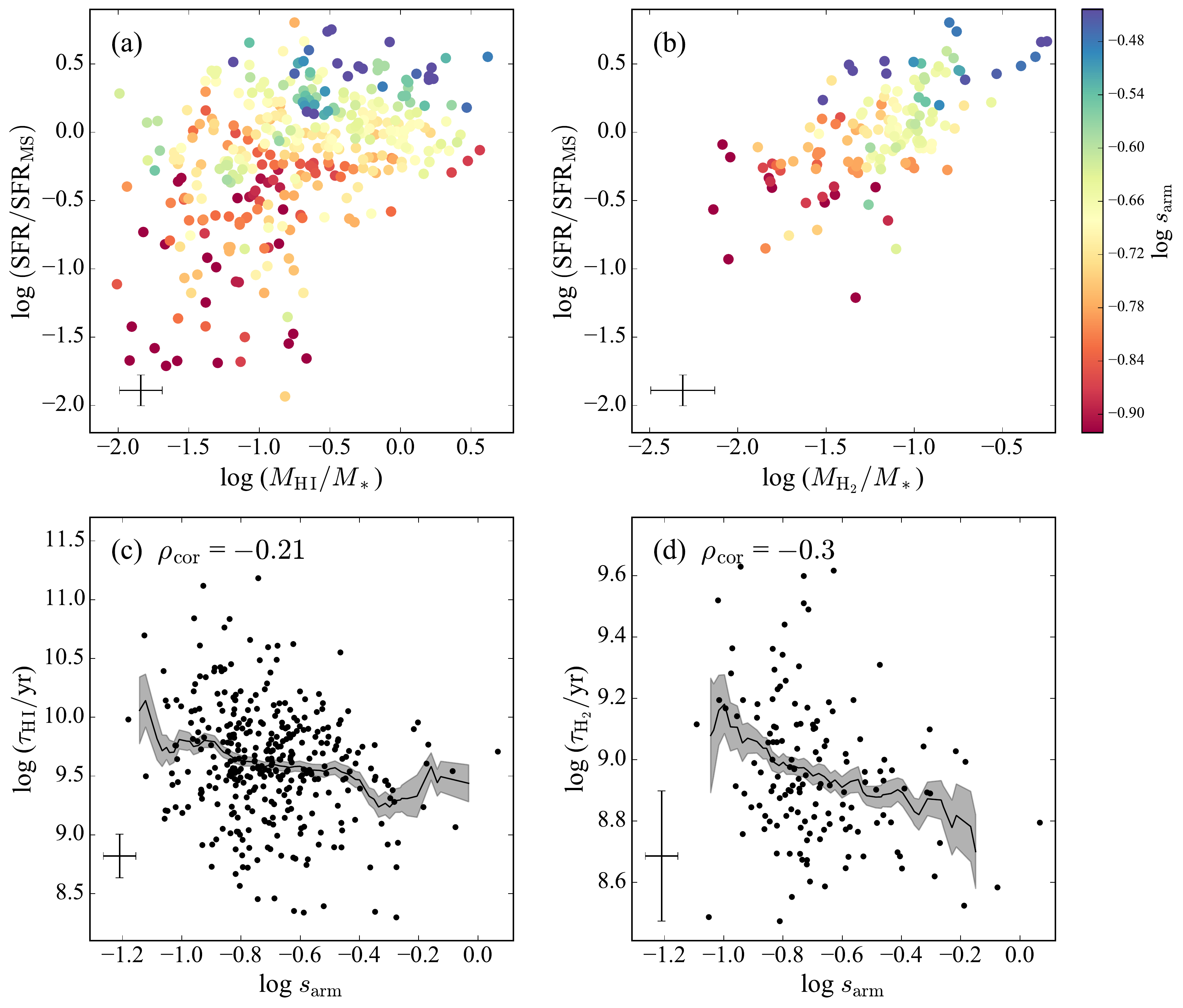}
\caption{ 
The connection between spiral arms and global star formation efficiency. Panels (a) and (b) examine the influence of spiral arm strength ($\log\,s_{\rm arm}$) on the relation between the distance to the MS [$\log\,({\rm SFR}/{\rm SFR}_{\rm MS})$] and the atomic gas mass fraction (\MHI$/M_*$) and molecular gas mass fraction ($M_{\rm H_2}/M_*$).  The color associated with each data point encodes the average arm strength of surrounding galaxies with $\Delta x \le 0.1$\,dex and $\Delta y \le 0.1$\,dex.  Panels (c) and (d) study the dependence of \HI\ ($\tau_{\rm H I}$) and \H2\ ($\tau_{\rm H_2}$) gas depletion timescale on $\log\,s_{\rm arm}$; the correlation coefficient $\rho_{\rm cor}$ is given.  Solid curves mark the mean value, while shaded regions mark the error of the mean value. The bars indicate typical measurement uncertainties.
}
\label{default}
\end{center}
\end{figure*}

\subsection{Spiral Arms Shorten Gas Depletion Timescale}

Star formation on global and sub-galactic scales couples more directly with the molecular gas than with the neutral atomic hydrogen \citep{Kennicutt1989, Leroy2008,Leroy2013, Schruba2011}. Figures~6a and 6b illustrate the well-known tendency for SFR to scale with gas fraction, especially in molecular form \citep{Saintonge2012, Saintonge2016, Genzel2015}.  Including the spiral arm measurements (given by the color coding) uncovers a second parameter that contributes to the scatter of $\log\,({\rm SFR}/{\rm SFR}_{\rm MS})$ at a given gas fraction.  Consistent with the above partial correlation analysis, at a fixed gas fraction (\MHI$/M_*$ or $M_{\rm H_2}/M_*$),  $\log\,({\rm SFR}/{\rm SFR}_{\rm MS})$ is higher for galaxies with more pronounced spiral arms.  The SFE, which describes how efficiently stars form from the gas reservoir, is the inverse of the gas depletion time ($\tau$), the time required for present-day star formation to consume the available gas supply.  Figures~6c and 6d show that spiral structure helps to boost the SFE or shorten the atomic ($\tau_{\rm H I}$; $\rho_{\rm cor}=-0.21$ with $p$ value $<0.001$) and molecular ($\tau_{\rm H_2}$; $\rho_{\rm cor}=-0.30$ with $p$ value $<0.001$) gas depletion time. Although gas depletion time also correlates with stellar mass surface density \citep{Huang2015}, we verify that the dependence on arm strength is not driven by $\mu_*$.  A partial correlation analysis between arm strength and \HI\ and \H2\ depletion time with the effect of $\mu_*$ removed yields a correlation coefficient of $\rho_{\rm cor}=-0.25$ and $\rho_{\rm cor}=-0.22$ at a statistically significant level ($p$ value $<0.001$ and $=0.009$, respectively).

\subsection{Connection between Spiral Arms and Star Formation}

The influence of spirals on star formation is implicit in the Silk-Elmegreen relation \citep{Elmegreen1997, Silk1997}, which posits that multiple passages of gas clouds through regions of deep gravitational potential of spiral waves favor the growth of gas clouds through cloud-cloud collision and their subsequent collapse.  A natural corollary is that during each passage through the spiral potential, the deeper potential generated by stronger arms would introduce more significant dynamical effects on the gas, thereby promoting the formation of stars. Besides, stronger spiral structure subjects the gas to more severe shock interaction \citep{Roberts1969}, elevates surface density along the arms \citep{Shetty2006, Dobbs2008}, drives gas inflows to create dense inner gaseous disks \citep{Kim2014}, and hence induces gravitational instability and more efficient star formation.
According to numerical simulations \citep{Dobbs2011, Kim2020}, stronger arms, be they density waves or recurrent material \citep{Pettitt2020}, enhance the SFR by a factor of $\sim 2$ (i.e., $0.3$\,dex), consistent with the result in Figure~6c and 6d.

These theoretical predictions can be tested through observations of nearby galaxies, although the results thus far have been mixed, even conflicting.  For example, the giant molecular clouds in the arm regions of M51 are brighter than those in the inter-arm regions of the galaxy, and their mass spectrum  presents a higher number density and more massive clouds, suggesting that the spiral arms promote cloud formation \citep{Colombo2014}.  However, M51 exhibits a large scatter in SFE along its spiral arms \citep{Meidt2013}.  The enhancement on SFE for the giant molecular clouds located on the spiral arms of NGC~6946 is more definitive \citep{Rebolledo2012}, but the same does not seem to hold for NGC~628, whose SFE is similar in its arm and inter-arm regions \citep{Foyle2010,Kreckel2016}.  The spiral arms of the Milky Way also appear to have little or no effect on the SFE \citep{Eden2015,Ragan2016}, despite the evidence that the spirals affect the distribution and kinematics of the dense gas \citep{Sakai2015}. This complexity may be traced to the fact that although shocks raise the gas density they also introduce turbulent and streaming motions, which prevent cloud collapse and curtail SFE \citep{Meidt2013, Leroy2017}.  Arm regularity, the key attribute of grand-design spirals, may be less important for triggering star formation than shock strength \citep{Elmegreen1986, Stark1987}, which is reflected in the arm strength. Within this backdrop, our results, firmly rooted on a robust statistical analysis of an unprecedentedly large number of nearby galaxies (more than 2 orders of magnitude larger than that of \citealt{Kendall2015}), strongly support the notion that spiral arm strength on average increases the global SFE on galactic scales. This study for the first time observationally addresses the role that spiral arms play in the general evolution of the galaxy population, by using a clean sample selection and usage of modern tools of understanding galaxy evolution (i.e., galaxy main sequence, deep surveys of gas masses).

A spiral galaxy needs external gas replenishment to maintain its star formation; otherwise, the galaxy consumes all the gas on a timescale much shorter than a Hubble time \citep[e.g.,][]{Bigiel2008}. One of the channels for gas replenishment is through accretion of halo gas onto the disk \citep{Fraternali2008}. An indirect, cyclic interaction may exist between spiral arms and the hot halo gas. Strong spirals trigger star formation along the arms \citep{Roberts1969}, supernovae and stellar winds from the massive stars drive a galactic fountain that mixes with the hot, low-metallicity halo and accelerates the cooling of the hot gas \citep{Fraternali2008}. The affected hot halo gas cools and rains down onto the disk to supplement the cold gas from which stars form.

We note, in passing, that the influence of spiral arm strength on the position of galaxies in the SFR-$M_*$ diagram naturally explains the weak dependence of the best-fitted MS on Hubble type. Late-type spirals lie slightly above early-type spirals on the SFR-$M_*$ diagram \citep{Catalan-Torrecilla2017,Canodiaz2019}. As late-type spirals on average have stronger arms than early-type spirals \citep{YuHo2020}, the influence of arm strength on SFR contributes to the observed dependence of MS position on Hubble type.

\subsection{The Effect of Measurement Uncertainty}

We use the spiral arm strength ($\log\,s_{\rm arm}$) to quantify the amplitude ($s_{\rm arm}$) of the spirals relative to that of the underlying axisymmetric disk.  A possible worry is contamination in the $r$ band by H$\alpha$ emission from star-forming regions along the spiral arms. It is known, for instance, that the relative arm amplitude measured in bluer bands is systematically stronger than that measured at longer wavelengths \citep{Yu2018a}. Images taken at 3.6\,$\mu$m, such as those in the Spitzer Survey of Stellar Structure in Galaxies (S$^4$G; \citealt{Sheth2010}), better trace the old stellar population, but the lack of overlap between our SDSS sample and S$^4$G precludes a direct comparison.  As the $r$ band is close to $R$, we perform a test using $R$-band images from the Carnegie-Irvine Galaxy Survey (CGS; \citealt{Ho2011}) of 108 spiral galaxies \citep{Yu2018a} that overlap with S$^4$G. A direct comparison of the spiral arm strength computed at 3.6\,$\mu$m and {\it R} band shows that they are strongly correlated (Pearson correlation coefficient $r=0.98$). Interestingly, the arm strength in the {\it R}\ band is actually $\sim$\,0.04\,dex weaker ($\sim$10\% lower in the relative arm amplitude),  not stronger, than the arm strength in the 3.6\,$\mu$m band, likely because of  dust emission along the spiral arms in the latter \citep{Kendall2011}.  Likewise, it has also been found that the arm strength in the $V$ band is on average slightly weaker than that in the 3.6\,$\mu$m band \citep{Kendall2015}. \cite{Querejeta2015} derive maps of the emission from old stars with dust emission removed from the S$^4$G 3.6\,$\mu$m images based on the method of \cite{Meidt2012}. We further use these maps to measure arm strength, which are found consistent with that in the {\it R} band.  The relative arm amplitude in the 3.6\,$\mu$m band is only 0.1\% higher than that in the {\it R}\ band.  The Pearson correlation coefficient between the arm strength computed from the two bands is $r=0.94$, with a median difference of only 0.0004\,dex. As the {\it i}\ band avoids both strong emission lines from star-forming regions and warm dust emission, we performed an additional test using a randomly selected subset of 200 SDSS galaxies from our sample, applying to the $i$ band the same data reduction procedures as used in the $r$ band.  The correlation between the {\it r}-band and {\it i}-band spiral arm strength is equally tight (Pearson correlation coefficient $r=0.98$), with the {\it r}-band strength only $\sim$\,0.015\,dex stronger ($\sim$\,3\% higher in the relative amplitude) than the {\it i}-band strength.  Any possible contamination from young stars is much smaller than the dynamical range of the arm strength itself.  We conclude that our main results based on $r$-band images are not affected significantly by H$\alpha$ emission from recent star formation.

To understand the influence of the radial variation of spiral strength on our results, we split the radial range of spiral arms we previously determined equally into an inner and an outer part. Figure~7a presents the correlation between spiral arm strength and the distance of galaxies from the MS, shown separately for different radial ranges.  We find no obvious difference among the results calculated in different radial ranges. We conclude that our main results are not sensitive to the choice of radial region used to compute the spiral arm strength. 

We then assess the effect of measurement uncertainty on SFR and stellar mass.
In addition to the spectral energy distribution (SED) fitting, the SFR can be estimated from other tracers such as the infrared luminosity or H$\alpha$ emission. These different SFR estimators agree well with each other for MS galaxies \citep{Salim2016}. However, determining the SFR for galaxies that lie below the MS [$\log$\,(sSFR$/$yr$^{-1}$) $< -11$] is challenging using any method. According to  \cite{Salim2016}, the mean random error of SFRs based on SED-fitting increases from 0.06\,dex at $\log$\,(sSFR$/$yr$^{-1}$) $= -9$, to 0.27\,dex at $\log$\,(sSFR$/$yr$^{-1}$) $= -11$, to 0.7\,dex at $\log$\,(sSFR$/$yr$^{-1}$) $= -12$.  As the majority of the quenched or nearly quiescent disk galaxies have intrinsically faint arms (Figure 2), the large random error on SFR is not expected to significantly affect our results for galaxies with low sSFR or low $\log\,({\rm SFR}/{\rm SFR}_{\rm MS})$. The stellar masses determined based on the photometry in the optical bands may be uncertain due to the lack of near-infrared data, which better trace the old stellar population, although stellar masses based on SED-fitting with and without near-infrared data agree  to within 0.1\,dex \citep{Muzzin2009}. To assess the influence of the random error from SFR and $M_*$, we perform bootstrapping calculations by first replacing each data point with a random number drawn from a normal distribution centered on the measured SFR or $M_*$ having a standard deviation equal to the error mentioned above, recalculating $\log\,({\rm SFR}/{\rm SFR}_{\rm MS})$, and then averaging the correlation between $\log\,({\rm SFR}/{\rm SFR}_{\rm MS})$ and spiral strength. We repeat this process 200 times and plot all the correlations in Figure~7b, delineated by the shaded region. We note that combining the original uncertainty and with the uncertainty from the bootstrapping process actually increases the uncertainty by $\sim$\,1.4 times, making the resulting shaded region slightly shallower than the original curve. However, overall, the random error from SFR or $M_*$ does not significantly change our results.

\begin{figure*}
\begin{center}
\includegraphics[width=0.8\textwidth]{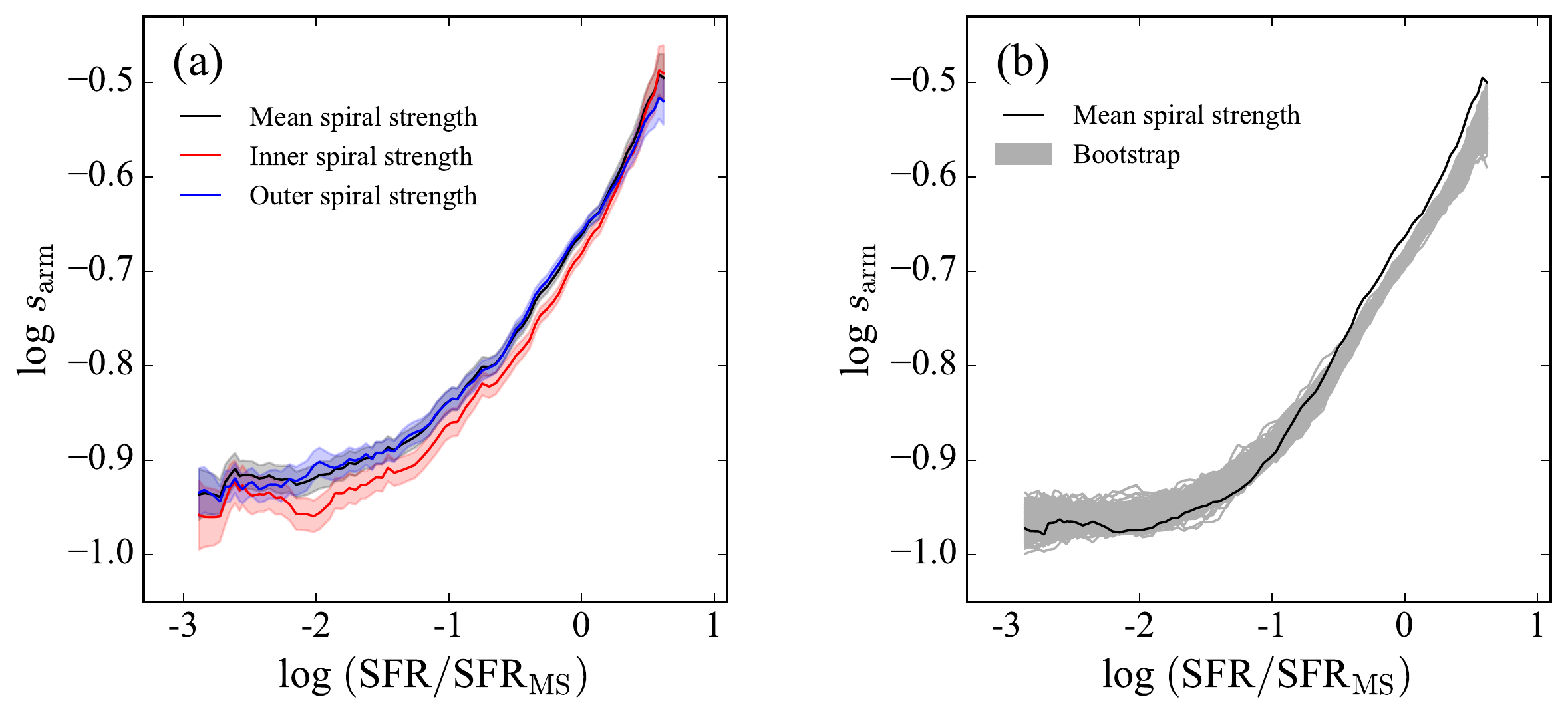}
\caption{
The effect of spiral strength ($\log s_{\rm arm}$) on the distance from the MS [$\log ({\rm SFR/SFR_{\rm MS})}$].  Panel (a) illustrates the arm strength over the full region (black), inner region (red), and outer region (blue); the shaded regions mark the error of the mean.  Panel (b) gives the mean relation between $\log\,s_{\rm arm}$ and $\log ({\rm SFR/SFR_{\rm MS})}$, along with the random error (grey shaded region) generated from 200 bootstrap iterations that include the random error of SFR and stellar mass.
} 
\label{default}
\end{center}
\end{figure*}

\section{Summary}

We study the impact of spiral structure on star formation by examining the relationships between the strength of spiral structure, SFR, and stellar mass using a large, statistically representative sample of 2226 nearby disk (S0 and spiral) galaxies.  We find that spiral arms are stronger above the star-forming galaxy MS and weaker below it, such that spiral arm strength increases with higher $\log\,({\rm SFR}/{\rm SFR}_{\rm MS})$, where ${\rm SFR_{MS}}$ denotes the SFR along the MS. Likewise, specific SFR increases with stronger spiral structure. This finding is confirmed for the full sample of 4378 disk galaxies using the $u-r$ color index, showing that the position of spiral galaxies on the blue cloud depends systematically on their arm strength. The dependence of $\log\,({\rm SFR}/{\rm SFR}_{\rm MS})$ on arm strength is independent of other galaxy structural parameters, such as bar strength, global stellar mass concentration (related to bulge dominance), and stellar mass surface density.  For the subset of galaxies with cold gas measurements, we find that spiral arm strength positively correlates with \HI\ and \H2\ mass fraction, even after removing the mutual dependence of these quantities on $\log\,({\rm SFR}/{\rm SFR}_{\rm MS})$.  We argue that the sensitivity of arm strength to gas content is in line with the notion that spiral arms are maintained by dynamical cooling provided by gas damping. We further show that stronger arms lead to higher $\log\,({\rm SFR}/{\rm SFR}_{\rm MS})$ for a given gas fraction. This results in a trend of increasing arm strength with shorter gas depletion time, independent of stellar mass surface density. These correlations suggest that spiral arms enhance star formation efficiency, and that the relationship between spiral arms and star formation is driven only in part by gas fraction. 

Spiral arms and gas content play an intertwined role in disk galaxies.  On the one hand, dissipation by gas damping maintains spiral structure, and on the other hand, spiral structure helps boost the star formation efficiency of the cold gas reservoir. This accounts for systematic variation of spiral arm strength on the star-forming MS.

\acknowledgments{We are grateful to the anonymous referee for helpful feedback. This work was supported by the National Science Foundation of China (11721303, 11991052) and the National Key R\&D Program of China
(2016YFA0400702).}

\end{document}